\documentclass[preprint,aps]{revtex4}
\usepackage{rotating}
\usepackage{graphicx}
\usepackage{dcolumn}
\usepackage{bm}
\usepackage{epsfig}
\usepackage{amsmath}
\usepackage{subfigure}

\def\be{\begin{equation}}
\def\lan{\left\langle}
\def\ran{\right\rangle}
\def\ee{\end{equation}}
\def\barr{\begin{array}}
\def\earr{\end{array}}

\def\l{\left}
\def\r{\right}
\def\dis{\displaystyle}
\def\ed{\end{document}}

\def\co{{\cal O}}

\begin{document}
\title{Shape effects on double beta decay of $^{70}$Zn and $^{150}$Nd in
deformed shell model}
\author{R. Sahu$^1$}
\email{rankasahu@rediffmail.com}
\author{V.K.B. Kota$^{2,3}$}
\affiliation{$^1$Physics Department, Berhampur University, Berhampur-760 007,
Odisha, INDIA}
\affiliation{$^2$Physical Research Laboratory, Ahmedabad - 380 009, INDIA \\
$^3$Department of Physics, Laurentian  University, Sudbury, ON P3E
2C6, CANADA }

\begin{abstract}

Using deformed shell model based on Hartree-Fock intrinsic states with $^{70}$Zn
as an example and  employing two realistic effective interactions, namely jj44b
that produces deformed shape and JUN45 that generates spherical shape, it is
demonstrated that the $0\nu$  and $2\nu$ nuclear transition matrix elements for
double beta decay will reduce considerably as we go from spherical shapes to
deformed shapes. This result is further substantiated by using $^{150}$Nd
as another example.

\end{abstract}

\maketitle

\section{Introduction}

Double beta decay (DBD) is an important open problem of current interest. There
are  many experimental efforts \cite{Kam, Exo, Gerda} and also theoretical
studies ( see for example \cite{IBM2,JS1} and references cited therein) in
particular for neutrinoless double beta decay. The nuclear transition matrix
elements (NTME)   $M^{0\nu}$ for neutrinoless double beta decay
($0\nu\beta\beta$)  are found to vary considerably in different nuclear models.
For example, QRPA values are always larger than the shell model values.
Most variants of QRPA assume the nucleus to be spherical and hence they do not
take into account the deformation of the nucleus. On the other hand, shell model
takes into account all possible  correlations, within the defined space, through
the effective interaction.  There are many studies which show that quadrupole
deformation reduces the $M^{0\nu}$. Some of them are as follows. (i) energy
density functional method including deformation and pairing fluctuations within
the finite range density dependent Gogny force \cite{Vaquero} where the  authors
show that deformation effects vary from $11$\% in $^{82}$Se to as high as $38$\%
in $^{128}$Te. (ii) The proton-neutron QRPA with a realistic residual
interaction \cite{Fang}. (iii) Nuclear shell model \cite{Menendez} where a 
quadrupole-quadrupole term is added to the effective interaction and  the
deformation  effects were studied by varying the strength of the quadrupole
term. In addition, there is also an attempt to study the effect of hexadecapole
deformation \cite{Chandra}. In this present report, we study the effect of
changing from spherical to deformed shape on $M^{0\nu}$ and $2\nu$ half-lives,
using realistic interactions within the deformed shell model (DSM) based on
Hartree-Fock  (HF) states. In the recent past, we have used DSM in the study DBD
(both $2\nu$ and $0\nu$) in a number of nuclei in the A=$60-90$ region
\cite{ge76,SK-kr,SK-se,SK-sr,SSK2,SSK,sk}. 

In Section 2 we will discuss briefly the formulas for $0\nu$ and $2\nu$
half-lives and the DSM model. Sections 3 and 4 give the results obtained using 
spherical and deformed shapes for $^{70}$Zn and $^{150}$Nd nuclei respectively. 
Section 5 gives conclusions.

\section{Double beta decay half-lives and deformed shell model} 

Half-life for  $0\nu$DBD for the 0$^+_i$ ground state (gs) of a initial
even-even nucleus decaying to the 0$^+_f$ gs of the final even-even nucleus is
given by \cite{IBM2}
\be
\l[ T_{1/2}^{0\nu}(0^+_i \to 0^+_f) \r]^{-1} =  G^{0\nu}\; \l(g_A\r)^4
\l| M^{0\nu} (0^+)\r|^2 \l(\dis\frac{\lan m_\nu \ran}{m_e}\r)^2 \;,
\label{eq.dbd1}
\ee
where $\lan m_\nu \ran$ is the effective neutrino mass and $G^{0\nu}$ is the
phase space integral (kinematical factor). The NTME $M^{0\nu}$ is essentially a
sum of a Gamow-Teller like ($M_{GT}$) and Fermi like ($M_F$)  two-body operators
(we are neglecting other parts as in \cite{SSK}). Then, 
\be
M^{0\nu} (0^+) = M^{0\nu}_{GT} (0^+) - \dis\frac{g_V^2}{g_A^2} 
M^{0\nu}_{F} (0^+) = \lan 0^+_f \mid\mid \co(2:0\nu) \mid\mid 0^+_i \ran \;.
\label{eq.dbd2}
\ee
As seen from Eq. (\ref{eq.dbd2}),  $0\nu$DBD half-lives are generated by the
two-body transition operator $\co(2:0\nu)$. The $g_A$ and $g_V$ are the weak
axial-vector and vector coupling constants. In our DSM analysis, we will use
$\co(2:0\nu)$ as given in \cite{SSK,sk} and it contains the neutrino potential,
Jastrow factor and a scale factor three.  It is important to mention that the
results in the present paper are  independent of the scale factor used in
\cite{SSK,sk} as we are  studying only the ratio of $M^{0\nu}$ for spherical and
deformed shapes. Eqs. (\ref{eq.dbd1}) and (\ref{eq.dbd2}) are a result of the
closure  approximation and this is justified {\it only if the exchange of the
light  Majorana neutrino is indeed the mechanism responsible for the
$0\nu$DBD}. 

For $2\nu$ DBD, half-life for the $0_I^{+} \rightarrow 0_F^{+}$ decay  
is given by \cite{Vogel}
\be
\l[ T_{1/2}^{2\nu}\r] ^{-1} = G_{2\nu } \;\l|M_{2\nu }\r|^2 \;.
\label{eqn19}
\ee
The kinematical factor $G_{2\nu} $ is independent of nuclear structure and the 
$M_{2\nu}$ is NTME given by,
\begin{equation} 
M_{2\nu}=\;\dis\sum_N\;\dis\frac{\lan 0_F^+ \mid\mid
\sigma\, \tau ^{+} \mid\mid 1_N^+ \ran \lan 1_N^+ \mid\mid \sigma\,\tau
^{+} \mid\mid 0_I^+ \ran}{\l[ E_N - (E_I + E_F)/2\r]/m_e} \;.
\label{eqn20}
\end{equation}
Here, $\l| 0_{I}^{+}\ran$, $\l| 0_{F}^{+}\ran$ and  $\l| 1_{N}^{+}\ran$ are the
initial, final and virtual intermediate states respectively and $E_{N}$ are the
energies of the intermediate nucleus. Similarly, $E_I$ and  $E_F$ are
the ground state energies of the parent and daughter nuclei (see 
\cite{SK-se,SSK} for further details).

We will use DSM to calculate $T_{1/2}^{2\nu}$ using the known values for
$G_{2\nu}$ and the NTME $M^{0\nu}$ for nuclei with close to spherical shape and
with deformation. In DSM, for a given nucleus, starting with a model space
consisting of a given set of single particle (sp) orbitals and effective
two-body Hamiltonian, the lowest energy intrinsic states are obtained by solving
the Hartree-Fock (HF) single particle equation self-consistently. Excited
intrinsic configurations are obtained by making particle-hole excitations over
the lowest intrinsic state.  These intrinsic states do not have definite angular
momenta and hence states of good angular momentum are  projected from them.
These good angular momentum states projected from different intrinsic states are
not in general orthogonal to each other. Hence they are orthonormalized and then
band mixing calculations are performed. DSM is well established to be a
successful model for transitional nuclei (with A=60-90) when sufficiently large
number of intrinsic states are included in the band mixing calculations. For
details see \cite{SSK} and references there in.

\section{Results for spherical and deformed shapes for $^{70}$Zn}

In the DSM calculations for $0\nu$ and $2\nu$ NTME for $^{70}$Zn, used is the
model space consisting of the orbitals  $^{2}p_{3/2}$, $^{1}f_{5/2}$,
$^{2}p_{1/2}$ and $^1g_{9/2}$ with $^{56}$Ni  as the core. It is important to
add that the same model space was used in many recent shell model and 
interacting boson model calculations \cite{IBM2,sm-ikeda,IBM2a} for $^{76}$Ge
and $^{82}$Se nuclei and $^{70}$Zn is in the same region. Within this space we
have employed two effective interactions. These are the jj44b interaction given
in \cite{jj44b} with spherical single particle energies for the four orbits
being $-9.6566$, $-9.2859$, $-8.2695$ and $-5.8944$ MeV respectively and  the
JUN45 interaction given in \cite{jun45} with spherical single particle energies
for the four orbits being $-9.8280$, $-8.7087$, $-7.8388$ and $-6.2617$ MeV
respectively. With these, we have first carried out axially symmetric  HF
calculations for the parent and the daughter nuclei $^{70}$Zn and $^{70}$Ge (for
$2\nu$DBD, also for the intermediate odd-odd nucleus $^{70}$Ga) for each of the
two effective interactions.

\begin{figure}
\includegraphics[width=6in]{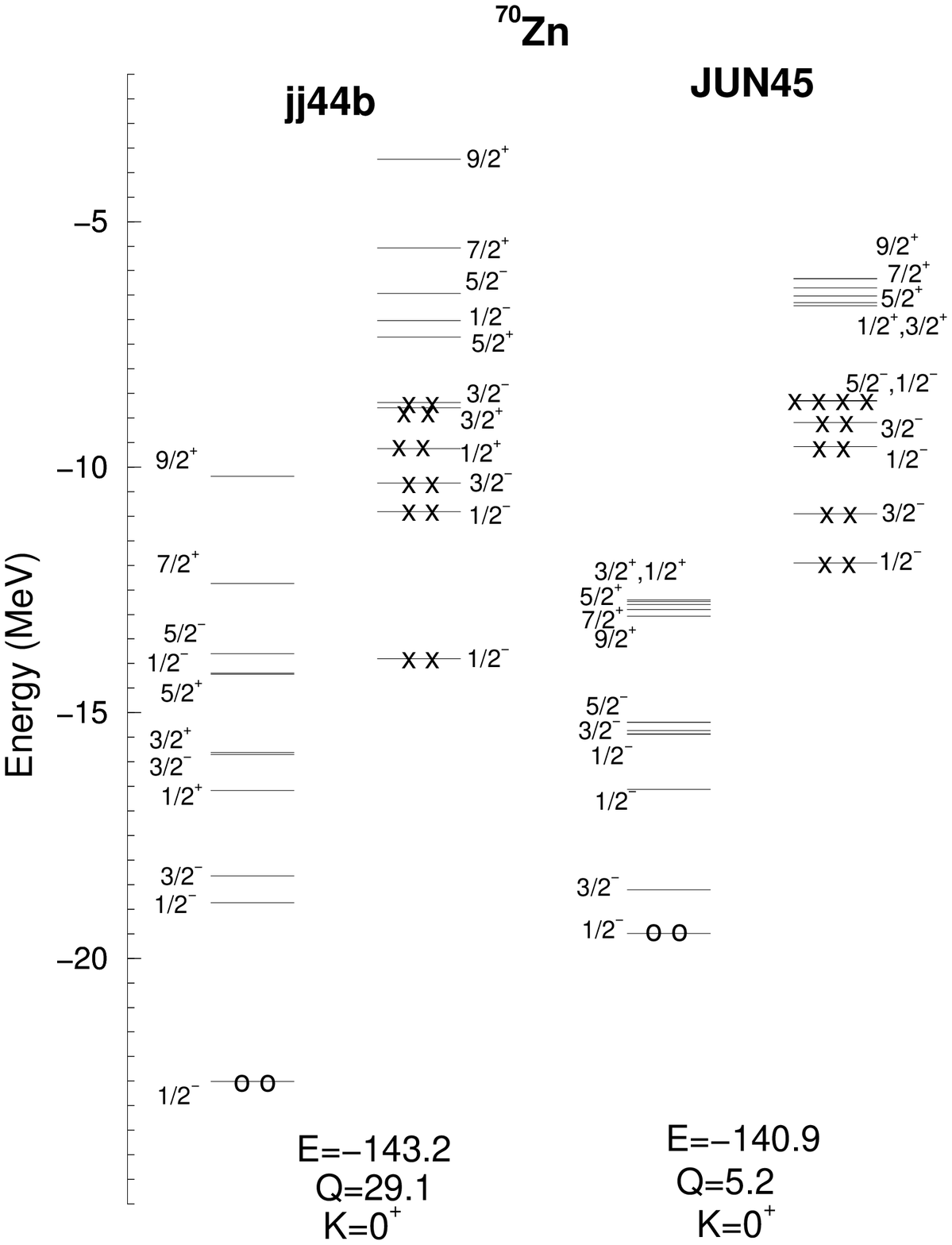}

\caption{HF single particle spectra for $^{70}$Zn  corresponding to lowest
prolate configurations generated by jj44b and JUN45  interactions.   In the
figures circles represent protons and crosses represent neutrons. The
Hartree-Fock energy  ($E$) in MeV, mass quadrupole moment ($Q$) in units of the
square of the oscillator length parameter and the total $K$ quantum number of
the lowest intrinsic states are given in the figure. Each occupied single
particle orbital is two fold degenerate because of time reversal symmetry.}

\label{fig1}
\end{figure}

\begin{figure}
\includegraphics[width=6in]{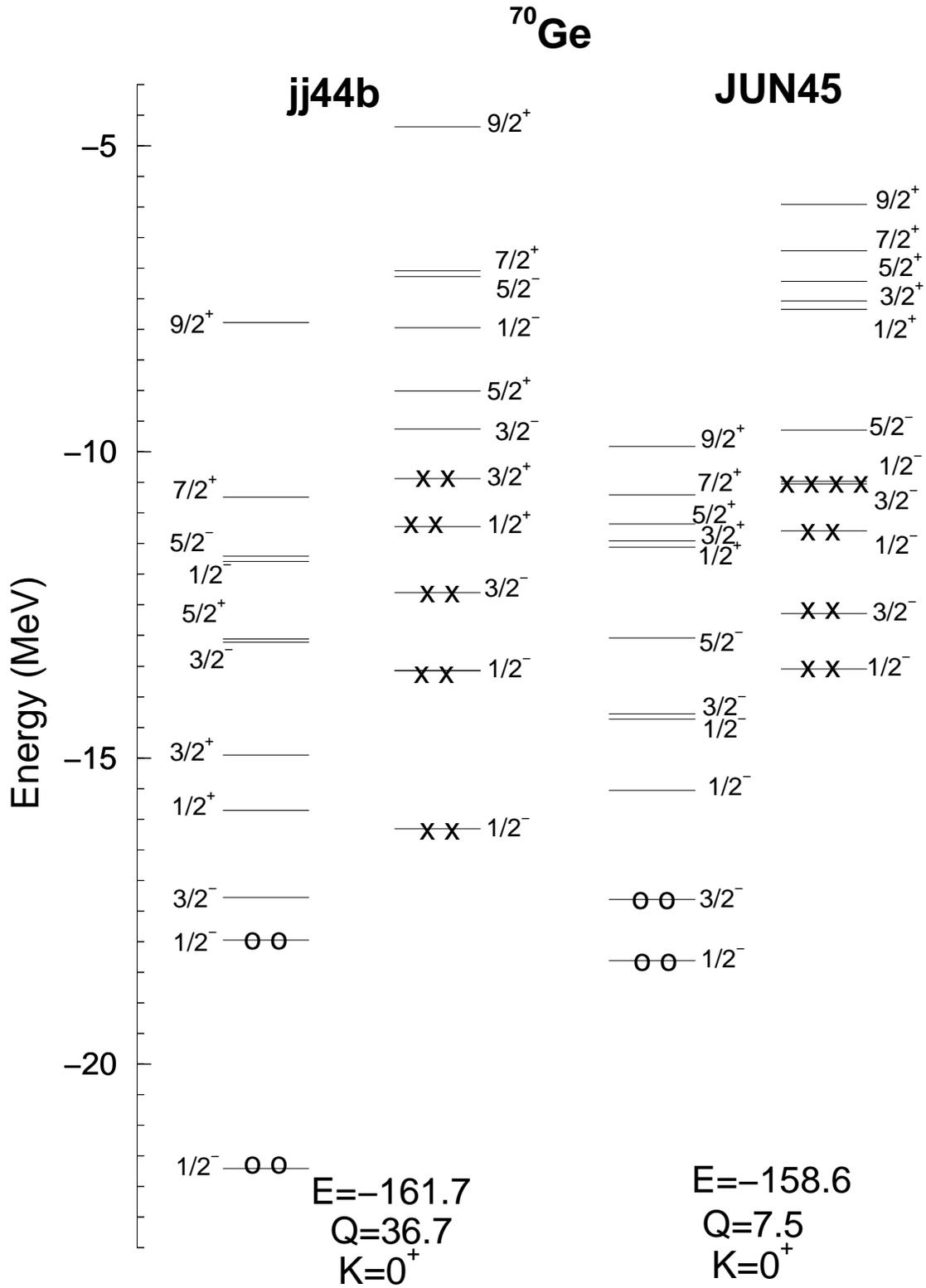}
\caption{Same as Fig. 1 but for $^{70}$Ge.}  
\label{fig2}
\end{figure}

The lowest energy HF solutions obtained using jj44b and JUN45 interactions  for
$^{70}$Zn are shown in  Fig. \ref{fig1} and similarly for $^{70}$Ge in Fig.
\ref{fig2}. It is clearly seen from the HF spectra in Figs. 1 and 2 that $jj44b$
generates deformed solutions for the ground state of $^{70}$Zn and $^{70}$Ge.
Note that the mass quadruple moment given in the figure gives a measure of 
deformation. With jj44b interaction, the DSM calculated energy  spectra for
these two nuclei are compressed compared to the experimental  spectrum. For
example, for $^{70}$Zn the calculated $E(2^+_1)$ is 0.355 MeV compared to  the
experimental value 0.885 MeV  and similarly for $^{70}$Ge they are 0.64 MeV and
1.04 MeV respectively \cite{sk}. But, at the same time the calculated $B(E2:
2^+_1 \rightarrow 0^+_1)$ (14 W.u. and 18.8 W.u. for $^{70}$Zn and $^{70}$Ga
respectively) are in  close agreement with data (16.5 W.u. and 20.9 W.u.
respectively). These results are reported in \cite{sk}. On the other hand,  as
seen from Figs. 1 and 2, the JUN45 interaction generates nearly spherical
solutions for these two nuclei. For this interaction, the spectrum is in close
agreement with data but the  $B(E2: 2^+_1 \rightarrow 0^+_1)$ is much smaller
than data value; see Fig. 3. Both the calculated energy spectra and $B(E2)$
values show that jj44b generates deformed shape and JUN45 generates spherical
shape. Therefore, calculations of NTME for $0\nu$ and $2\nu$ using these
interactions will give information about shape effects on NTME.

\subsection{$M^{0\nu}$ and $T^{2\nu}_{1/2}$ for $^{70}$Zn}

In the DSM calculations of $0\nu$ NTME using jj44b interaction, we have
considered 30 lowest intrinsic  states with   $K=0^+$ for $^{70}$Zn, 26 lowest
intrinsic configurations with $K = 0^+$    for the daughter nucleus $^{70}$Ge
(up to 6 MeV excitation in both nuclei) by making particle-hole  excitations
over the lowest HF intrinsic states generated for these nuclei. We project out
$0^+$ states from each of these intrinsic states and  then perform  band mixing
calculations as discussed above. Similarly for JUN45 we have considered 81
lowest intrinsic  states with   $K=0^+$ for $^{70}$Zn, 68 lowest intrinsic
configurations with $K = 0^+$ for the daughter nucleus $^{70}$Ge (up to 5 MeV
excitation in both nuclei) by making particle-hole  excitations over the lowest
HF intrinsic states generated for these nuclei. With jj44b interaction, the NTME $M^{0\nu}$
comes out to be 2.13. However, for JUN45 interaction the value is 4.56. Thus we
see a reduction by a factor $2$ in $M^{0\nu}$ as we go from spherical to deformed
shape. Thus DSM is consistent with other theoretical calculations mentioned in
Section 1. For further confirmation of this result we will turn $2\nu$ NTME.

For $2\nu$ DBD, firstly as discussed in \cite{sk}, $[ E_N - (E_I + E_F)/2 ] = 
[1.1537 + E_{1^+}]$ MeV and DSM is used to calculate $E_{1^+}$ for the
intermediate nucleus $^{70}$Ga as well as the numerator in the sum in Eq.
(\ref{eqn20}). In the DSM calculations using jj44b and JUN45 interactions,  we
have considered the same intrinsic states for $^{70}$Zn and $^{70}$Ge as in  the
calculation of $0\nu$ NTME given above. In addition, for the intermediate
nucleus  $^{70}$Ga, 65 intrinsic states with $K = 1^+$ or  $K = 0^+$ are used
for jj44b interaction and 114 intrinsic states with JUN45 interaction. These are
generated by making particle-hole  excitations over the lowest HF intrinsic
state generated for this nucleus. We project out $1^+$ states from each of these
intrinsic states and  then perform a band mixing calculation as discussed above.
Taking the  phase space factor  $0.32 \times 10^{-21}$ yr$^{-1}$ given in
\cite{Vogel}, the DSM value for $T^{2\nu}_{1/2}$ is $3.39 \times 10^{23}$ yr for
jj44b interaction and $4.94 \times 10^{22}$ yr for JUN45 interaction. Thus, for
$2\nu$ DBD also the NTME (proportional to inverse of the half-life) is  reduced
by a factor of $\sim 2.6$ as we go from spherical (JUN45) to deformed (jj44b)
shape. 

\section{Results for spherical and deformed shapes for $^{150}$Nd}

We then considered the double beta decay of $^{150}$Nd as another example.  In
the calculation of $^{150}$Nd and the daughter nucleus $^{150}$Sm, we used
modified Kuo-Herling interaction \cite{Chou}  with a Z = 50, N = 82 core. The
protons are in the $^3s_{1/2}$, $^2d_{3/2}$, $^2d_{5/2}$,  $^1g_{7/2}$ and 
$^1h_{11/2}$ orbits with spherical single particle energies -7.92783, -7.95519,
-8.67553, -9.59581 and -6.83792 MeV,  respectively. The  neutrons are in the 
$^3p_{1/2}$, $^3p_{3/2}$, $^2f_{5/2}$,  $^2f_{7/2}$, $^1h_{9/2}$ and 
$^1i_{13/2}$ orbits with spherical single particle energies for these orbits
-1.050, -1.625, -1.964, -2.380, -0.895 and -0.261 MeV,  respectively. The model
space used here is same as in the recent interacting  boson model calculations
\cite{IBM2,IBM2a}. As described before, we obtain the lowest energy HF
configuration by solving the axially symmetric HF equation self-considently. The
lowest energy HF solutions for $^{150}$Nd and $^{150}$Sm  obtained with this
interaction are shown in figures  \ref{nd150hf} and \ref{sm150hf}. As seen from
the quadrupole moments, both the nuclei have large deformations. To generate HF
intrinsic states with nearly spherical shape, we artificially decreased the
effective interation to 1/10th of its value and perform axially symmetric HF
calculation as described above. These nearly spherical HF single particle states
are shown in figures \ref{nd150hf} and \ref{sm150hf}. As expected, these two
solutions have small quadrupole moment and hence small deformation. We then
proceed to calculate $0\nu$ NTME with the deformed and spherical solutions,
separately. For the deformed solutions, we considered 38 intrinsic states with
$K=0^+$  for $^{150}$Nd. Similarly we consider 17 intrinsic states with $K=0^+$
for the daughter nucleus $^{150}$Sm. Now, the NTME $M^{0\nu}$ comes out to be
1.35. Then in the second case where we reduced the effective interaction matrix
elements by a factor of 10, we considered 62 intrinsic states both for the
parent and the daughter nuclei. The NTME $M^{0\nu}$ comes out to be 3.06. Thus
in this  case also the $0\nu$ nuclear matrix element increases by a factor of
2.33 as  we go from deformed to spherical shape.

\section{Conclusions}

We have demonstrated using DSM model and $^{70}$Zn as the example and  employing
two realistic effective interactions, namely jj44b that produces deformed shape
and JUN45 that generates spherical shape, that the $0\nu$ and $2\nu$ NTME for
double beta decay will reduce considerably as we go from spherical shapes to
deformed shapes. This result is further confirmed by a calculation of $0\nu$
NTME for $^{150}$Nd. Our results reported in this paper are an important
addition, since the calculations are performed using two realistic interactions
proposed for the same model space and DSM is transperent in determining shapes,
to the existing analysis of deformation effects on NTME for DBD.

We would like to mention that since we have  not included all the spin-orbit
partners in the basis space,  the Ikeda sum rule is not satisfied in our
calculations.  However, as mentioned before, many recent  shell model and
interacting boson model calculations \cite{IBM2, sm-ikeda,IBM2a} used the same
model space. Violation of Ikeda sum rule may not be a serious deficiency in our
calculations since we are only studying the change in double beta decay matrix 
elements with the change of nuclear shape. Calculations using much larger set of
single particle orbits (these also need new effective interactions) are for 
future.

\begin{acknowledgments}

\noindent RS is thankful to DST (Government of India) for financial support. 

\end{acknowledgments}

\begin{figure}
\includegraphics[width=4in]{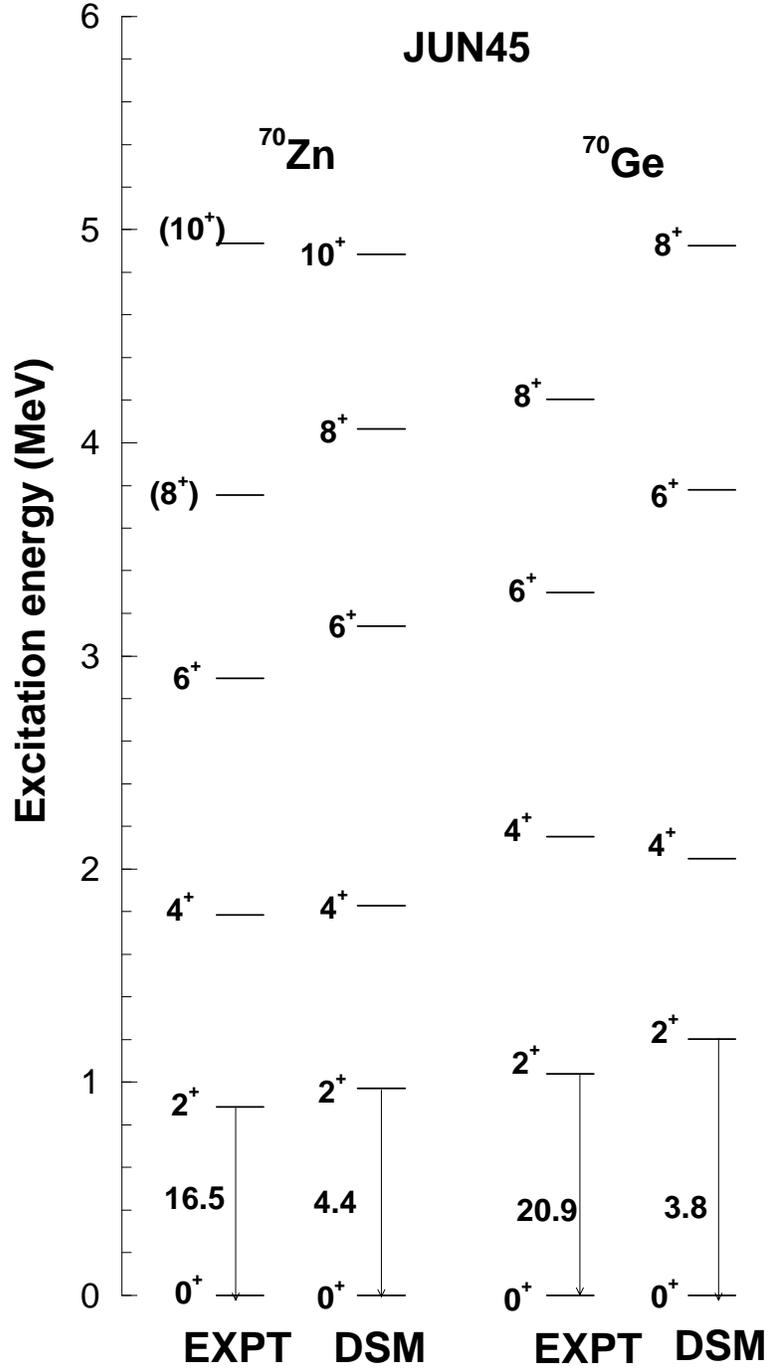}

\caption{The calculated yrast levels, using JUN45 interaction, for $^{70}$Zn  and
$^{70}$Ge compared with  experiment. The experimental data are from ref 
\cite{nndc}. The quantities near the arrows represent B(E2) values in W.u. and
the effective charges used are same as in \cite{sk}.}
\label{fig3}
\end{figure}

\begin{figure}
\includegraphics[width=6in]{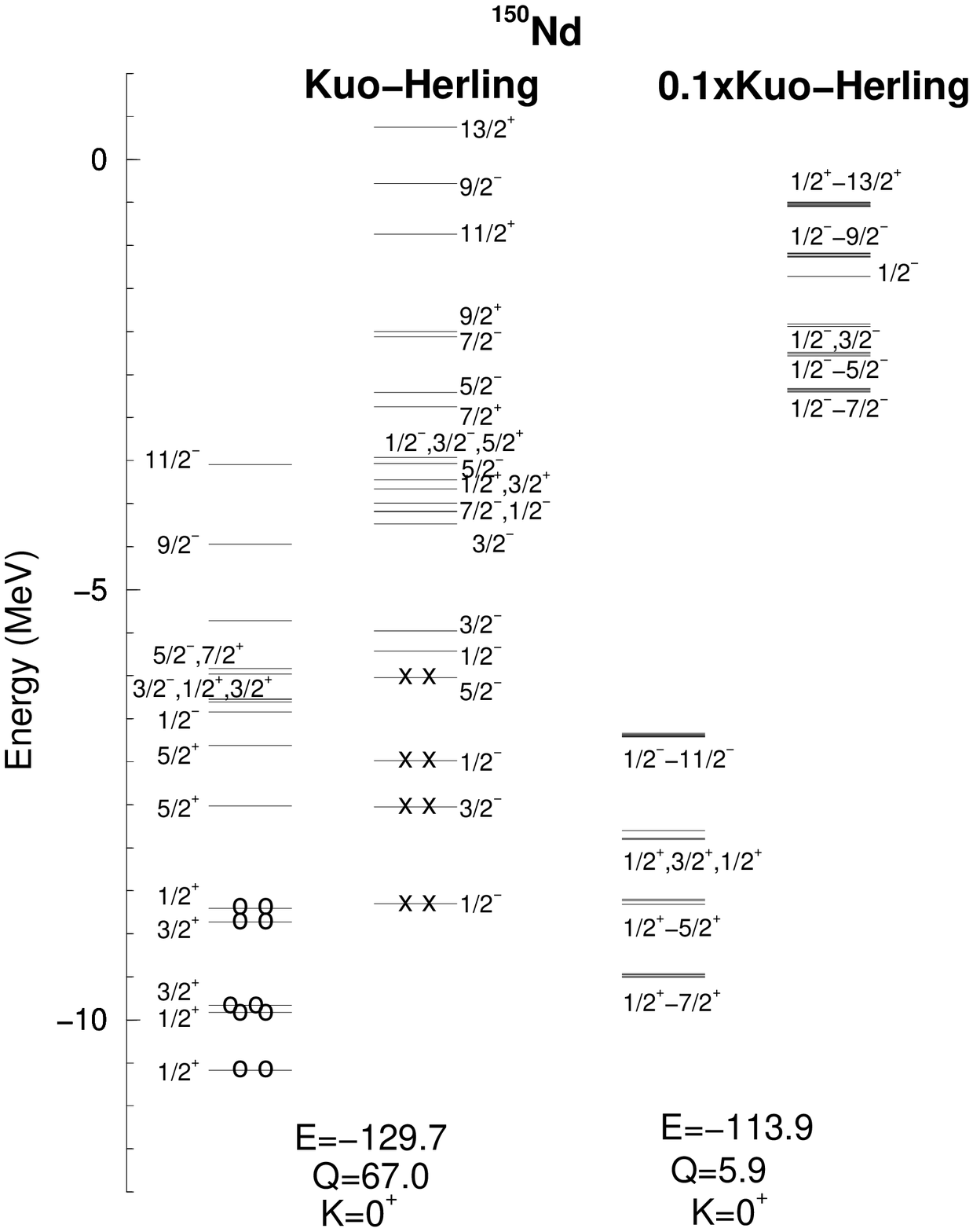}
\caption{Same as in Fig. 1 but for $^{150}$Nd with Kuo-Herling interaction
(see text for details). In the spherical solution, the levels are degenerate
as expected.}
\label{nd150hf}
\end{figure}

\begin{figure}
\includegraphics[width=6in]{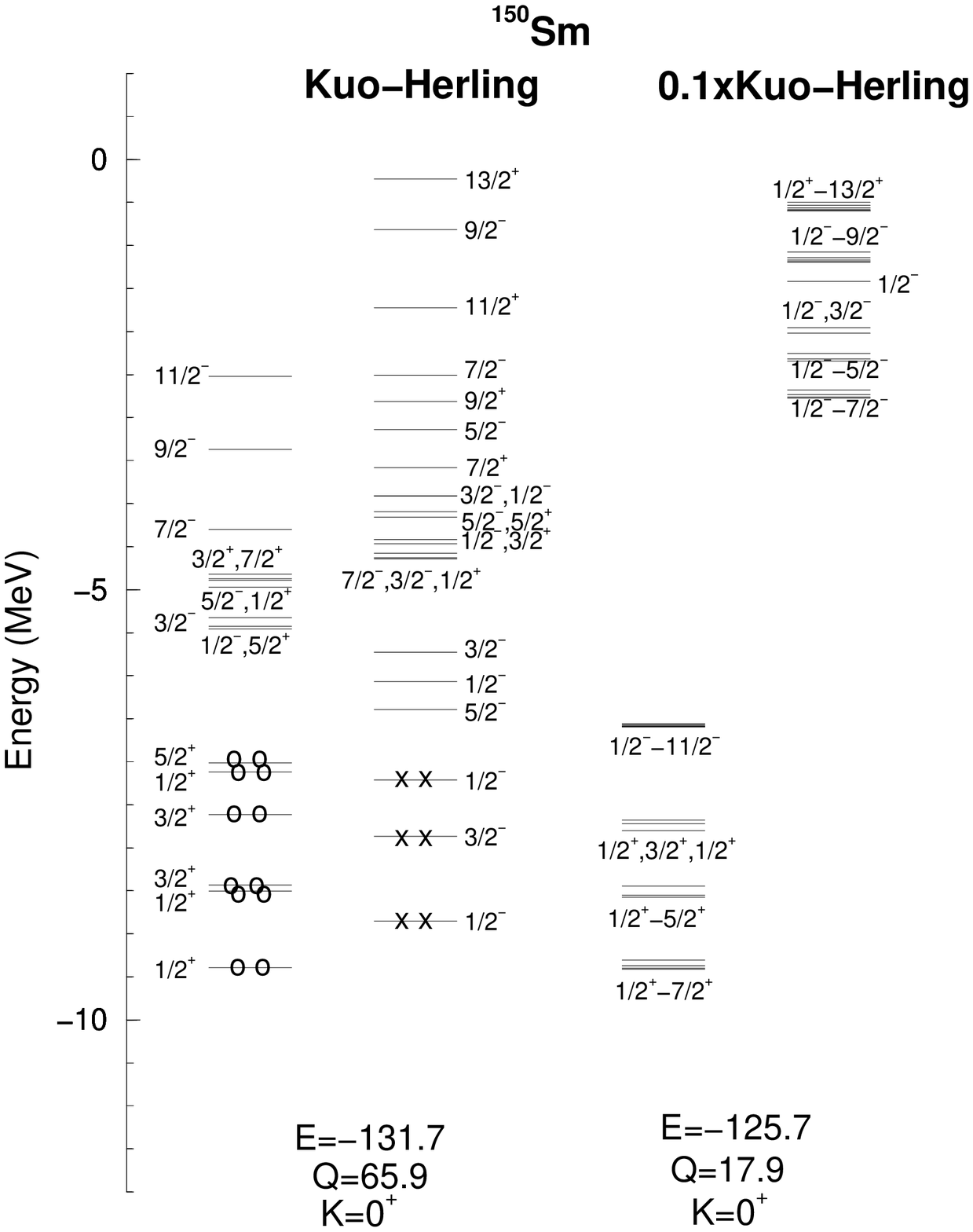}
\caption{Same as in Fig. 1 but for $^{150}$Sm with Kuo-Herling interaction
(see text for details). In the spherical solution, the levels are degenerate
as expected.}
\label{sm150hf}
\end{figure}
\end{document}